\documentclass{article}%
\usepackage{amssymb}
\usepackage{amsmath}
\usepackage{amsfonts}
\usepackage{graphicx}%
\providecommand{\U}[1]{\protect\rule{.1in}{.1in}}
\begin{document}

\title{Influence of initial correlations on evolution of correlation function of \\a subsystem interacting with a quantum field (heat bath) and polaron\\mobility}
\author{Victor F. Los\\Institute for Magnetism, 36-b Vernadsky Blvd., 03142, Kiev, Ukraine}

\begin{abstract}
A regular approach to accounting for initial correlations, which allows to go
beyond the unrealistic random phase (initial product state) approximation in
deriving the evolution equations, is suggested..An exact homogeneous equation
for a two-time equilibrium correlation function for the dynamical variables of
a subsystem interacting with a boson field (heat bath) is obtained. No
conventional approximation like RPA or Bogoliubov's principle of weakening of
initial correlations is used. The obtained equation takes into account the
initial correlations in the kernel governing its evolution. The solution to
this equation is found In the second order of the kernel expansion in the
electron-phonon interaction, which demonstrates that generally the initial
correlations influence the correlation function's evolution in time. It is
shown that this influence vanishes on a large timescale. The developed
approach is applied to the Fr\"{o}hlich polaron and the low-temperature
polaron mobility (which was under a long-time debate) is found with a
correction due to initial correlations .

\ \ \ \ \ \ \ \ \ \ \ \ \ \ \ \ \ \ \ \ \ \ \ \ \ \ \ \ \ \ \ \ \ \ \ \ \ \ \ \ \ \ \ \ \ \ \ PACS
number(s): 05.30.-d, 71.38.-k

\end{abstract}
\maketitle

\section{Introduction}

Strict derivation of kinetic (irreversible) equations from the underlying
reversible microscopic dynamics of the many-particle systems remains an
important task of statistical physics due to its fundamental importance and
various applications of these kinetic equations to the specific physical
problems. Actually, to achieve this goal, several assumptions are
conventionally used. The principal ones include the factorizing initial
conditions (random phase approximation or "molecular chaos") corresponding to
uncorrelated initial state and Bogoliubov's principle of weakening of initial
correlations \cite{Bogoliubov}. Such approximations, which, in fact, suggest
that all initial (at the initial moment of time $t_{0}$) correlations vanish
on a large timescale, allow for obtaining the kinetic equations valid only on
a large timescale. Thus, the main problem is to include initial correlations
into consideration in order to obtain the equations which describe the system
evolution on an arbitrary timescale and will, in particular, enable studying
the influence of initial correlations on the relaxation process. The method
which allows to include initial correlations into the kernel governing the
evolution of the relevant part of the statistical operator of the many-body
system (like gas or liquid) and obtain homogeneous generalized master
equations valid on any timescale was developed in works \cite{Los JPA, Los
JSP, Los TMP, Los Evolution Equations}.

Another important case is a subsystem $S$ of the total system interacting with
an environment (a thermal bath) $\Sigma$ of the remaining part of a whole
system. In this case, the factorizing initial condition for the statistical
operator of the whole system $\rho(t)$%
\begin{equation}
\rho(t_{0})=\rho_{S}(t_{0})\rho_{\Sigma} \label{0}%
\end{equation}
is often used ($\rho_{S}(t)=Tr_{\Sigma}\rho(t)$ is a subsystem statistical
operator defined as a trace over a thermal bath states and $\rho_{\Sigma}$ is
the heat bath statistical operator). Such an approximation is incorrect in
principle as it has been clearly pointed out by van Kampen \cite{van Kampen}.
It may, however, be expected, that according to Bogoliubov's principle of
weakening of initial correlations, at a large timescale%
\begin{equation}
t-t_{0}>>t_{cor}, \label{0a}%
\end{equation}
where $t_{cor\text{ }}$is the correlation time for fluctuations in a heat
bath, the initial correlations vanish and thus the condition (\ref{0}) can be used.

The important (for quantum field theory and condensed matter physics) example
of such a system is a particle interacting with a quantum field (heat bath).
The electron-phonon system in a crystal is one of the realizations of such a
system, which is important for practical applications. The notion of a polaron
(a quasiparticle made of an electron and a cloud of phonons) introduced by
Landau \cite{Landau} has become crucial for understanding a crystal
conductivity, especially in polar crystals \cite{Pekar}. Experimentally, the
notion of the polaron is important to the understanding of a wide variety of
materials, a few of which include the conductivity of semiconductors,
high-temperature superconductors and the BEC with an immersed impurity atom
(for more details see overview \cite{Devreese arXiv}). Recent experimental
advances in the field of ultra-cold atoms offer new possibilities for
experimental studies of polarons and have resulted in revived interest in the
polaron problem.

The polaron ground-state (free energy) calculation is now done quite
accurately over the complete interaction range \cite{Feynman 1955}. At the
same time, the Fr\"{o}hlich polaron dynamics has been a matter of debate over
a long time. Actually, three results on low-temperature polaron mobility have
been obtained. Mainly, the discrepancy between the Kadanoff \cite{Kadanoff}
and the Feynman-Hellwarth-Iddings-Platzman (FHIP) results \cite{FHIP} was
discussed and only recently it has been understood \cite{Sels and Brosens
2014}. There was also my result on polaron mobility, which is different from
the two mentioned results \cite{Los 1984}. It turned out, that neither the
Kadanoff result nor the FHIP one are correct. Instead, the polaron mobility
was found to agree with that of \cite{Los 1984} (see \cite{Sels and Brosens
2014} and for a detailed discussion \cite{Devreese arXiv}). It is worth
mentioning that the result of the paper \cite{Los 1984} was obtained in
\cite{Sels and Brosens 2014} independently and by using a method (truncation
of the Liouville equation for the reduced Wigner distribution function)
different from that of \cite{Los 1984}.

The polaron mobility can be calculated by different methods. The path integral
approach of FHIP and the relaxation time approximation for the Boltzmann
equation of Kadanoff were already mentioned. In \cite{Los 1984}, the
Green-Kubo formula was employed for calculation of the conductivity tensor.
The equilibrium current-current correlation function was calculated by the
Green superoperator technique in \cite{Los TMP 1979}. In the papers \cite{Los
and Martynenko 1983} and \cite{Los and Martynenko 1986}, the closed Markovian
equation for the density matrix of a subsystem interacting with a boson field
(serving as a thermostat) was derived using the same Green superoperator
method. As in the path integrals approach of FHIP, the bosonic degrees of
freedom were eliminated exactly from the obtained equation. This equation for
the density matrix was applied in \cite{Los and Martynenko 1986} to the
polaron problem. In the paper \cite{Los Wigner}, the closed equation for the
Wigner function of a subsystem interacting with a heat bath and under the
influence of inhomogeneous electric field was obtained and applied to the
polaron problem.

However, in the paper \cite{Los 1984}, the initial correlations were
disregarded for the purposes of calculating the current-current correlation
function, i.e., the condition like (\ref{0}) was used: the equilibrium
distribution function in this correlation function (which in this case plays
the role of the initial condition) $\rho(\beta)=(Tre^{-\beta H})^{-1}e^{-\beta
H}$ ($\beta=1/k_{B}T$, $H$ is the Hamiltonian of the whole electron+boson
field system) was factorized (the correlation between an electron and the
thermostat was omitted). The reason was that if we consider the long times
$t-t_{0}\thicksim\tau_{rel}>>\beta$ ($\hbar=1$, $\tau_{rel}$ is a subsystem
relaxation time, $t_{cor}\thicksim\beta$), then the influence of initial
correlations on an evolution process is small or vanishes (see also \cite{van
Kampen}). The same approximation was also used in papers \cite{Los and
Martynenko 1983, Los and Martynenko 1986, Los Wigner}, where the initial
condition for the statistical operator of the whole system at the initial
moment of time was selected (conventionally) as the factorized one (the
product of the statistical operators of the subsystem and the thermal bath).

In this paper, by applying the projection operator technique, we obtain the
exact closed homogeneous equation for the relevant part of the correlation
function of a subsystem interacting with a boson field (thermostat). It
includes the contribution of initial correlations ($\rho(\beta)$ is accounted
for in its exact form) to the kernel governing the evolution of a correlation
function. The solution to this equation in the second order approximation in
the electron-phonon interaction is found. It demonstrates that initial
correlations influence the correlation function evolution in time on an
arbitrary timescale. We show that this influence on the time dependence of the
correlation function vanishes on a large timescale. The obtained equation is
applied to the polaron problem and the low-temperature polaron mobility is
calculated. In particular, small contribution of initial correlations to the
result of \cite{Los 1984} is found.

\section{Exact equation for a subsystem correlation function accounting for
initial correlations}

We consider a subsystem $S$ which interacts with a large equilibrium system
(the heat bath $\Sigma$). The Hamiltonian of the whole system $S+\Sigma$ is
\begin{equation}
H=H_{S}+H_{\Sigma}+H_{i}, \label{1}%
\end{equation}
where $H_{S}$, $H_{\Sigma}$ and $H_{i}$ are the Hamiltonians of the subsystem,
the Bose field (heat bath) and the subsystem-Bose field interaction, respectively.

The two-time correlation function for the arbitrary operators $A$ and $B$
related to the subsystem $S$ interacting with a heat bath $\Sigma$ can be
written as
\begin{equation}
<A_{S}(t)B_{S}(0)>=Tr_{S,\Sigma}[B_{S}(0)G_{S}(t,\beta)].\label{2}%
\end{equation}
Here,%
\begin{equation}
G_{S}(t,\beta)=\rho(\beta)e^{iLt}A_{S}(0),\label{3}%
\end{equation}
$Tr_{S,\Sigma}$ means taking a trace over the states of the whole system
$S+\Sigma$, the averaging is performed with the equilibrium distribution of
the whole system $\rho(\beta)=(Tre^{-\beta H})e^{-\beta H}$, $L$ is the
Liouville superoperator acting on the arbitrary operator $D$ according to the
rules%
\begin{equation}
LD=[H,D]=HD-DH,e^{iLt}D=e^{iHt}De^{-iHt},\hbar=1,\label{4}%
\end{equation}

and $L=L_{S}+L_{\Sigma}+L_{i}$ in accordance with the Hamiltonian (\ref{1}).

We will consider the dynamics of the correlation function (\ref{1}) by
splitting the function (\ref{3}) into the relevant $R_{S}(t,\beta)$ and
irrelevant $I_{S}(t,\beta)$ parts
\begin{equation}
R_{S}(t,\beta)=PG_{S}(t,\beta),I_{S}(t,\beta)=QG_{S}(t,\beta) \label{5}%
\end{equation}
by means of the projection operator \ \
\begin{align}
P(...)  &  =\rho_{\Sigma}(\beta)Tr_{\Sigma}(...),Q=1-P,P^{2}=P,Q^{2}%
=Q,PQ=0,\nonumber\\
\rho_{\Sigma}(\beta)  &  =(Tr_{\Sigma}e^{-\beta H_{\Sigma}})e^{-\beta
H_{\Sigma}}, \label{6}%
\end{align}
where $Tr_{\Sigma}$ means performing the trace over the states of the heat
bath $\Sigma$ and $\rho_{\Sigma}(\beta)$ is the heat bath equilibrium distribution.

It is not difficult to see that the dynamics of the correlation function
(\ref{2}) is completely defined by the relevant part of $G_{S}(t,\beta)$, i.e.%
\begin{equation}
<A_{S}(t)B_{S}(0)>=Tr_{S,\Sigma}[B_{S}(0)R_{S}(t,\beta)]. \label{7}%
\end{equation}
Thus, we need an evolution equation for the function $R_{S}(t,\beta)$. Using
the identity $P+Q=1$, (\ref{3}) and (\ref{5}), we have%
\begin{align}
\frac{\partial}{\partial t}R_{S}(t,\beta)  &  =iPL[R_{S}(t,\beta
)+I_{S}(t,\beta)],\nonumber\\
\frac{\partial}{\partial t}I_{S}(t,\beta)  &  =iQL[R_{S}(t,\beta
)+I_{S}(t,\beta)]. \label{8}%
\end{align}
The solution to the equation (\ref{8}) for a complementary (irrelevant)
function $I_{S}(t,\beta)$ reads%
\begin{align}
I_{S}(t,\beta)  &  =i%
%TCIMACRO{\dint \limits_{0}^{t-t_{0}}}%
%BeginExpansion
{\displaystyle\int\limits_{0}^{t-t_{0}}}
%EndExpansion
M_{Q}(\tau,\beta)QLR_{S}(t-\tau,\beta)d\tau+M_{Q}(t-t_{0},\beta)I_{S}%
(t_{0},\beta),\nonumber\\
M_{Q}(t,\beta)  &  =\exp(iQLQt). \label{9}%
\end{align}
Substituting this solution to the first equation of (\ref{8}) and selecting
the initial moment of time $t_{0}=0$, we obtain the following equation for the
relevant function $R_{S}(t,\beta)$%
\begin{align}
\frac{\partial}{\partial t}R_{S}(t,\beta)  &  =iPLPR_{S}(t,\beta)-PLQ%
%TCIMACRO{\dint \limits_{0}^{t}}%
%BeginExpansion
{\displaystyle\int\limits_{0}^{t}}
%EndExpansion
M_{Q}(\tau,\beta)QLR_{S}(t-\tau,\beta)d\tau\nonumber\\
&  +iPLQM_{Q}(t,\beta)I_{S}(0,\beta). \label{10}%
\end{align}
This equation is closed but contains the inhomogeneous (last in (\ref{10}))
term accounting for correlation between the subsystem $S$ and the thermostat
$\Sigma$ at the initial moment of time. These initial correlations, as it
follows from the definition (\ref{5}), are given by%
\begin{align}
Q\rho(\beta)  &  =\rho(\beta)-P\rho(\beta)=\rho(\beta)-\rho_{\Sigma}%
(\beta)\rho_{S}(\beta),\nonumber\\
\rho_{S}(\beta)  &  =Tr_{\Sigma}\rho(\beta), \label{10a}%
\end{align}
where $\rho_{S}(\beta)$ is is the reduced distribution function for a
subsystem. If we disregard these initial correlations by approximating the
distribution function $\rho(\beta)$ in (\ref{3}) as $\rho(\beta)=\rho
_{S}(\beta)\rho_{\Sigma}(\beta)$, then the inhomogeneous term in (\ref{10})
$I_{S}(0,\beta)=0$. Such a factorization of the distribution function for a
system under consideration was used in our earlier papers (mentioned in the
Introduction), particularly in \cite{Los 1984}. In this paper, we will not use
this approximation and include the initial correlations into consideration.

Direct accounting for initial correlations by means of solving the
inhomogeneous equation (\ref{10}) for $R_{S}(t,\beta)$ is not effective and
may lead to the appearance of the growing with time terms in this solution
(see, e.g. \cite{Bogoliubov}). Instead, we will include the initial
correlations term into the kernel governing the evolution in time of function
$R_{S}(t,\beta)$ and thus obtain the homogeneous evolution equation for this
function. Moreover, the expansion of this equation kernel in a small parameter
is, as a rule, more effective than the expansion of an equation itself. This
approach is in line with that employed in works \cite{Los JPA, Los JSP, Los
TMP, Los Evolution Equations}. However, the essential difference with these
works is that in our case we have the explicit expression for initial
correlations defined by the equilibrium distribution $\rho(\beta)$ for the
whole system $S+\Sigma$.

Thus, to convert Eq. (\ref{10}) into the exact homogeneous equation, we use
the following identity%
\begin{equation}
e^{-\beta H}=e^{-\beta H_{0}}-%
%TCIMACRO{\dint \limits_{0}^{\beta}}%
%BeginExpansion
{\displaystyle\int\limits_{0}^{\beta}}
%EndExpansion
d\lambda e^{-\lambda H_{0}}H_{i}e^{\lambda H}e^{-\beta H}, \label{11}%
\end{equation}
where $H_{0}=H_{S}+H_{\Sigma}$. Using (\ref{11}) and taking into account that
$Qe^{-\beta H_{0}}=0$, $P+Q=1$, $e^{iLt}e^{-iLt}=1$, the irrelevant initial
correlation term $I_{S}(0,\beta)=QG_{S}(0,\beta)=Q\rho(\beta)A_{S}(0)$ can be
then presented as%
\begin{align}
I_{S}(0,\beta)  &  =-C_{Q}(t,\beta)[R_{S}(t,\beta)+I_{S}(t,\beta)],\nonumber\\
C_{Q}(t,\beta)  &  =%
%TCIMACRO{\dint \limits_{0}^{\beta}}%
%BeginExpansion
{\displaystyle\int\limits_{0}^{\beta}}
%EndExpansion
d\lambda Qe^{-\lambda H_{0}}H_{i}e^{\lambda H}e^{-iLt}. \label{12}%
\end{align}
As a result, we have two coupled equations (\ref{9}) and (\ref{12}) for
$I_{S}(t,\beta)$ and $I_{S}(0,\beta)$. Inserting the solution of these
equations for the irrelevant initial correlations term $I_{S}(0,\beta)$ into
Eq. (\ref{10}), we obtain the exact homogeneous equation for the relevant
function $R_{S}(t,\beta)$ which we are looking for%
\begin{align}
\frac{\partial}{\partial t}R_{S}(t,\beta)  &  =iPL[1-K_{Q}(t,\beta
)][R_{S}(t,\beta)+i\int\limits_{0}^{t}d\tau M_{Q}(\tau,\beta)QLPR_{S}%
(t-\tau,\beta)],\nonumber\\
K_{Q}(t,\beta)  &  =M_{Q}(t,\beta)[1+C_{Q}(t,\beta)M_{Q}(t,\beta)]^{-1}%
C_{Q}(t,\beta). \label{13}%
\end{align}

Equation (\ref{13}) is the main result of this section. It differs from the
conventional Generalized Master Equations (see, e.g. \cite{Los TMP 1979}) for
the relevant part of the correlation function (or statistical operator) by the
modified time-local (first in the r.h.s. of (\ref{13})) term and non-Markovian
(integral) term which now contain the contribution of initial correlations.
This modification is defined by operator $K_{Q}(t,\beta)$. It is essential
that this operator is proportional to the interaction $H_{i}$, which makes it
conventional for expanding the kernel of Eq. (\ref{13}) in the subsystem-heat
bath interaction. If we put $K_{Q}(t,\beta)=0$ (disregard initial
correlations), we obtain the equation for the relevant part of the
distribution function (\ref{2}), which was used in \cite{Los 1984}.

\section{Electrons in a crystal interacting with bosons}

First of all we are interested in the problem of a polaron mobility in a polar
crystal. In this case,%
\begin{equation}
H_{S}=T(\mathbf{p}),H_{\Sigma}=\sum\limits_{\mathbf{k}}\omega_{\mathbf{k}%
}b_{\mathbf{k}}^{+}b_{\mathbf{k}}, \label{17}%
\end{equation}
where $T(\mathbf{p})$ is the kinetic energy of the electron with momentum
$\mathbf{p}$, $\omega_{\mathbf{k}}$ is the energy of the Boson field quantum
(phonon) with the wave vector $\mathbf{k}$ ($\hbar=1$), $b_{\mathbf{k}}^{+}$,
$b_{\mathbf{k}}$ are the Bose-operators of the creation and annihilation of
phonons, respectively. It is often sufficient to use the linear in
Bose-operators interaction Hamiltonian
\begin{equation}
H_{i}=\sum\limits_{\mathbf{k}}(V_{\mathbf{k}}e^{i\mathbf{kr}}b_{\mathbf{k}%
}+V_{\mathbf{k}}^{\ast}e^{-i\mathbf{kr}}b_{\mathbf{k}}^{+}), \label{18}%
\end{equation}
where $V_{\mathbf{k}}$ is an electron-phonon interaction coefficient and $r$
is an electron radius-vector.

It is easy to see that for projection operator $P$ (\ref{6}) commutes with
$L_{S}$, and%

\begin{equation}
PL_{S}Q=QL_{S}P=0,PL_{\Sigma}=L_{\Sigma}P=0,PL_{\Sigma}Q=QL_{\Sigma}P=0.
\label{19}%
\end{equation}
Moreover, for $H_{\Sigma}$ and $H_{i}$ defined by (\ref{17}) and (\ref{18}),%
\begin{equation}
PL_{i}P=0. \label{20}%
\end{equation}
The properties (\ref{19}), (\ref{20}) allow for simplification of our basic
equations (\ref{13}) and (\ref{16}).

The superoperators entering these equations can be expanded into the series
with respect to the interaction $H_{i}$. For example,
\begin{equation}
\lbrack1+C_{Q}(t,\beta)M_{Q}(t,\beta)]^{-1}=%
%TCIMACRO{\dsum \limits_{n=0}^{\infty}}%
%BeginExpansion
{\displaystyle\sum\limits_{n=0}^{\infty}}
%EndExpansion
(-1)^{n}\left[  C_{Q}(t,\beta)M_{Q}(t,\beta)\right]  ^{n}. \label{21}%
\end{equation}
The exponentials in the superoperators $C_{Q}(t,\beta)$ and $M_{Q}(t,\beta) $
can be also expanded in $H_{i}$ or $L_{i}$. At weak subsystem-thermostat
interaction, these series may be regarded as the perturbation theory expansions.

Taking into account (\ref{19}), (\ref{20}), we can rewrite Eq. (\ref{13}) as
\begin{align}
\frac{\partial}{\partial t}R_{S}(t,\beta)  &  =iPL_{S}R_{S}(t,\beta
)-iPL_{i}K_{Q}(t,\beta)R_{S}(t,\beta)\nonumber\\
&  -PL_{i}[1-K_{Q}(t,\beta)]\int\limits_{0}^{t}d\tau M_{Q}(\tau,\beta
)L_{i}PR_{S}(t-\tau,\beta)]. \label{21a}%
\end{align}
It is seen from (\ref{21a}), that in the zero approximation in the interaction
$H_{i}$,
\begin{equation}
R_{S}(t-\tau,\beta)=\exp[iL_{S}(t-\tau)]R_{S}(0,\beta)=\exp(-iL_{S}\tau
)R_{S}(t,\beta). \label{21b}%
\end{equation}
Assuming that the electron-phonon interaction is proportional to a small
parameter, taking into account (\ref{19}) - (\ref{21b}) and that
$C_{Q}(t,\beta)$ is proportional to $H_{i}$, we can convert Eq. (\ref{21a}) in
the following time-local equation in the second order in $H_{i}$ approximation
for the kernel governing the evolution of $R_{S}(t,\beta)$%

\begin{align}
\frac{\partial}{\partial t}R_{S}(t,\beta)  &  =[iPL_{S}P-PL_{i}%
%TCIMACRO{\dint \limits_{0}^{t}}%
%BeginExpansion
{\displaystyle\int\limits_{0}^{t}}
%EndExpansion
d\tau e^{iL_{0}\tau}L_{i}e^{-iL_{S}\tau}P\nonumber\\
&  -iPL_{i}e^{iL_{0}t}%
%TCIMACRO{\dint \limits_{0}^{\beta}}%
%BeginExpansion
{\displaystyle\int\limits_{0}^{\beta}}
%EndExpansion
d\lambda e^{-\lambda H_{0}}H_{i}e^{\lambda H_{0}}e^{-iL_{0}t}P]R_{S}(t,\beta)
\label{23}%
\end{align}
\qquad

The first two terms on the right-hand-side of Eq. (\ref{23}) coincide with
that of \cite{Los 1984} but in this paper we do not restrict ourselves (at
this stage) to the large times $t\backsim\tau_{rel}>>t_{0}$ (at \cite{Los
1984}, the upper limit of integration over $\tau$ was extended to infinity)
where $t_{0}=\max(t_{S},t_{\Sigma})$ ($t_{S}$ is the electron characteristic
collision time, e.g. of the $1/k_{B}T$ order; $t_{\Sigma}$ is of the order of
fluctuations correlation time in the heat bath, which is defined by the
characteristic phonon frequency). The third term in the r.h.s. of (\ref{22})
is absent in the corresponding equation of \cite{Los 1984}. It defines the
contribution of initial correlations which are treated here on the equal
footing with collisions (given by the second term).

Equation (\ref{23}) completely defines the evolution equation for the
correlation function (\ref{7}) in the second order in $H_{i}$ approximation
with accounting for initial correlations%
\begin{equation}
\frac{\partial}{\partial t}<A_{S}(t)B_{S}(0)>=Tr_{S,\Sigma}[B_{S}%
(0)\frac{\partial}{\partial t}R_{S}(t,\beta)]. \label{24}%
\end{equation}
Having in mind the further application of this equation to the polaron
mobility problem, let us take the subsystem operator $B_{S}(0)$ dependent only
on the momentum operator, $B_{S}(0)=B_{S}(\mathbf{p})$. Then, the trace over
the subsystem (electron states) in (\ref{24}) can be conveniently calculated
using the eigenfunctions of $H_{S}$ (\ref{17}), i.e., the eigenfunctions
$|\mathbf{p>}$ of the momentum operator. Therefore, we need to find
$\frac{\partial}{\partial t}R_{S}(t,\beta;\mathbf{p})$, i.e., the equation
(\ref{23}) for the diagonal matrix element of $<\mathbf{p}|R_{S}%
(t,\beta)|\mathbf{p}>=R_{S}(t,\beta;\mathbf{p})$.

It is not difficult to see that for $H_{S}=T(\mathbf{p})$ the diagonal matrix
element of the first term on the r.h.s. of (\ref{23}) $<\mathbf{p|}L_{S}%
R_{S}(t,\beta)|\mathbf{p}>=0$ (we used the rule (\ref{4})). Let us now
consider the contribution of electron collisions with the bath quanta
(phonons) to Eq. (\ref{23}) (the second term on the r.h.s. of (\ref{23})).
Applying consecutively the superoperators and projection operator $P$ in this
term to $R_{S}(t,\beta)$, making use of the rules (\ref{4}), and taking into
account that for the Hamiltonian, given by Eqs. (\ref{17}), (\ref{18}),
\begin{align}
e^{iH_{\Sigma}t}b_{\mathbf{k}}e^{-iH_{\Sigma}t}  &  =e^{-i\omega_{\mathbf{k}%
}t}b_{\mathbf{k}},e^{iH_{\Sigma}t}b_{\mathbf{k}}^{+}e^{-iH_{\Sigma}%
t}=e^{i\omega_{\mathbf{k}}t}b_{\mathbf{k}}^{+},\nonumber\\
&  <b_{\mathbf{k}}b_{\mathbf{k}_{1}}>_{\Sigma}=0,<b_{\mathbf{k}}%
^{+}b_{\mathbf{k}_{1}}^{+}>_{\Sigma}=0,\nonumber\\
&  <b_{\mathbf{k}}b_{\mathbf{k}_{1}}^{+}>_{\Sigma}=(1+N_{\mathbf{k}}%
)\delta_{\mathbf{kk}_{1}},<b_{\mathbf{k}}^{+}b_{\mathbf{k}_{1}}>_{\Sigma
}=N_{\mathbf{k}}\delta_{\mathbf{kk}_{1}},\nonumber\\
&  <...>_{\Sigma}=Tr_{\Sigma}(...\rho_{\Sigma}),N_{\mathbf{k}}=[\exp
(\beta\omega_{\mathbf{k}})-1]^{-1}, \label{25}%
\end{align}
we obtain the collision term in (\ref{23}) with
\begin{align}
&  <\mathbf{p|}PL_{i}%
%TCIMACRO{\dint \limits_{0}^{t}}%
%BeginExpansion
{\displaystyle\int\limits_{0}^{t}}
%EndExpansion
d\tau e^{iL_{0}\tau}L_{i}e^{-iL_{S}\tau}PR_{S}(t,\beta)|\mathbf{p}>\nonumber\\
&  =2\sum\limits_{\mathbf{k}}|V_{\mathbf{k}}|^{2}\{[\frac{\sin([T(\mathbf{p}%
-\mathbf{k})-T(\mathbf{p})+\omega_{\mathbf{k}}]t)}{T(\mathbf{p}-\mathbf{k}%
)-T(\mathbf{p})+\omega_{\mathbf{k}}}(1+N_{\mathbf{k}})\nonumber\\
&  +\frac{\sin([T(\mathbf{p}-\mathbf{k})-T(\mathbf{p})-\omega_{\mathbf{k}}%
]t)}{T(\mathbf{p}-\mathbf{k})-T(\mathbf{p})-\omega_{\mathbf{k}}}N_{\mathbf{k}%
}]R_{S}(t,\beta;\mathbf{p})\nonumber\\
&  -[\frac{\sin([T(\mathbf{p}-\mathbf{k})-T(\mathbf{p})-\omega_{\mathbf{k}%
}]t)}{T(\mathbf{p}-\mathbf{k})-T(\mathbf{p})-\omega_{\mathbf{k}}%
}(1+N_{\mathbf{k}})\nonumber\\
&  +\frac{\sin([T(\mathbf{p}-\mathbf{k})-T(\mathbf{p})+\omega_{\mathbf{k}}%
]t)}{T(\mathbf{p}-\mathbf{k})-T(\mathbf{p})+\omega_{\mathbf{k}}}N_{\mathbf{k}%
}]R_{S}(t,\beta;\mathbf{p-k})\}. \label{26}%
\end{align}
In order to obtain (\ref{26}), we also used that
\begin{equation}
<\mathbf{p}_{1}\mathbf{|}e^{\pm i\mathbf{kr}}|\mathbf{p}_{2}\mathbf{>=}%
\delta\mathbf{(p}_{2}\pm\mathbf{k}-\mathbf{p}_{1}\mathbf{).} \label{27}%
\end{equation}
Note, that Eq. (\ref{26}) is valid for arbitrary time $t$, in contrast to the
corresponding collision term obtained in \cite{Los 1984}.

The diagonal matrix element of the third term in the r.h.s. of Eq. (\ref{23}),
which defines the contribution of initial correlations, can be calculated in
the same way. The result is
\begin{align}
&  <\mathbf{p}|iPL_{i}e^{iL_{0}t}%
%TCIMACRO{\dint \limits_{0}^{\beta}}%
%BeginExpansion
{\displaystyle\int\limits_{0}^{\beta}}
%EndExpansion
d\lambda e^{-\lambda H_{0}}H_{i}e^{\lambda H_{0}}e^{-iL_{0}t}PR_{S}%
(t,\beta)|\mathbf{p}>\nonumber\\
&  =i\sum\limits_{\mathbf{k}}|V_{\mathbf{k}}|^{2}\int\limits_{0}^{\beta
}d\lambda\{[e^{i[T(\mathbf{p}-\mathbf{k})-T(\mathbf{p})+\omega_{\mathbf{k}%
}](t+i\lambda)}(1+N_{\mathbf{k}})\nonumber\\
&  +e^{i[T(\mathbf{p}-\mathbf{k})-T(\mathbf{p})-\omega_{\mathbf{k}%
}](t+i\lambda)}N_{\mathbf{k}}]R_{S}(t,\beta;\mathbf{p})\nonumber\\
&  -[e^{-i[T(\mathbf{p}-\mathbf{k})-T(\mathbf{p})-\omega_{\mathbf{k}%
}](t+i\lambda)}(1+N_{\mathbf{k}})\nonumber\\
&  +e^{-i[T(\mathbf{p}-\mathbf{k})-T(\mathbf{p})+\omega_{\mathbf{k}%
}](t+i\lambda)}N_{\mathbf{k}}]R_{S}(t,\beta;\mathbf{p}-\mathbf{k})\}
\label{28}%
\end{align}
The integration over $\lambda$ (which can be easily performed) is saved in
(\ref{28}) just for making this result more short in writing.

Inserting (\ref{26}) and (\ref{28}) in (\ref{24}), we arrive at the following
equation for the correlation function%
\begin{align}
\frac{\partial}{\partial t}  &  <A_{S}(t)B_{S}(0)>=Tr_{\Sigma}\int
d\mathbf{p}B_{S}(\mathbf{p})\frac{\partial}{\partial t}R_{S}(t,\beta
;\mathbf{p})\nonumber\\
&  =-Tr_{\Sigma}\int d\mathbf{p}B_{S}(\mathbf{p})[\gamma_{S}(t,\beta
;\mathbf{p})+\gamma_{S}^{\prime}(t,\beta;\mathbf{p})]R_{S}(t,\beta
;\mathbf{p}), \label{29}%
\end{align}
where
\begin{align}
\gamma_{S}(t,\beta;\mathbf{p})  &  =2\sum\limits_{\mathbf{k}}|V_{\mathbf{k}%
}|^{2}\frac{B_{S}(\mathbf{p})-B_{S}(\mathbf{p}-\mathbf{k})}{B_{S}(\mathbf{p}%
)}\nonumber\\
&  \times\{(1+N_{\mathbf{k}})\frac{\sin([T(\mathbf{p})-T(\mathbf{p}%
-\mathbf{k})-\omega_{\mathbf{k}}]t)}{T(\mathbf{p})-T(\mathbf{p}-\mathbf{k}%
)-\omega_{\mathbf{k}}}\nonumber\\
&  +N_{\mathbf{k}}\frac{\sin([T(\mathbf{p})-T(\mathbf{p}-\mathbf{k}%
)+\omega_{\mathbf{k}}]t)}{T(\mathbf{p})-T(\mathbf{p}-\mathbf{k})+\omega
_{\mathbf{k}}}\},\nonumber\\
\gamma_{S}^{\prime}(t,\beta;\mathbf{p})  &  =i\sum\limits_{\mathbf{k}%
}|V_{\mathbf{k}}|^{2}\frac{B_{S}(\mathbf{p})-B_{S}(\mathbf{p}-\mathbf{k}%
)}{B_{S}(\mathbf{p})}\nonumber\\
&  \times\{(1+N_{\mathbf{k}})e^{i[T(\mathbf{p}-\mathbf{k})-T(\mathbf{p}%
)+\omega_{\mathbf{k}}]t}\frac{1-e^{-[T(\mathbf{p}-\mathbf{k})-T(\mathbf{p}%
)+\omega_{\mathbf{k}}]\beta}}{T(\mathbf{p}-\mathbf{k})-T(\mathbf{p}%
)+\omega_{\mathbf{k}}}\nonumber\\
&  +N_{\mathbf{k}}e^{i[T(\mathbf{p}-\mathbf{k})-T(\mathbf{p})-\omega
_{\mathbf{k}}]t}\frac{1-e^{-[T(\mathbf{p}-\mathbf{k})-T(\mathbf{p}%
)-\omega_{\mathbf{k}}]\beta}}{T(\mathbf{p}-\mathbf{k})-T(\mathbf{p}%
)-\omega_{\mathbf{k}}}\}. \label{30}%
\end{align}
The term $\gamma_{S}^{\prime}(t,\beta;\mathbf{p})$ is due to initial
correlations and is valid, as well as $\gamma_{S}(t,\beta;\mathbf{p})$, for
all $t$ and temperature ($\beta^{-1}$).

Thus, we have the following equation for function $R_{S}(t,\beta;\mathbf{p})
$, which completely defines the correlation function (\ref{7}),%
\begin{equation}
\frac{\partial}{\partial t}\ln R_{S}(t,\beta;\mathbf{p})=-[\gamma_{S}%
(t,\beta;\mathbf{p})+\gamma_{S}^{\prime}(t,\beta;\mathbf{p})]. \label{31}%
\end{equation}
By integrating this equation over $t$ (taking into account (\ref{30})), we
obtain%
\begin{equation}
R_{S}(t,\beta;\mathbf{p})=R_{S}(0,\beta;\mathbf{p})\exp\{-[\Gamma_{S}%
(t,\beta;\mathbf{p})+\Gamma_{S}^{\prime}(t,\beta;\mathbf{p})]\}, \label{32}%
\end{equation}
where%
\begin{align}
\Gamma_{S}(t,\beta;\mathbf{p})  &  =\int\limits_{0}^{t}\gamma_{S}(\tau
,\beta;\mathbf{p})d\tau=2\sum\limits_{\mathbf{k}}|V_{\mathbf{k}}|^{2}%
\frac{B_{S}(\mathbf{p})-B_{S}(\mathbf{p}-\mathbf{k})}{B_{S}(\mathbf{p}%
)}\nonumber\\
&  \times\{(1+N_{\mathbf{k}})\frac{1-\cos([T(\mathbf{p})-T(\mathbf{p}%
-\mathbf{k})-\omega_{\mathbf{k}}]t)}{[T(\mathbf{p})-T(\mathbf{p}%
-\mathbf{k})-\omega_{\mathbf{k}}]^{2}}\nonumber\\
&  +N_{\mathbf{k}}\frac{1-\cos([T(\mathbf{p})-T(\mathbf{p}-\mathbf{k}%
)+\omega_{\mathbf{k}}]t)}{[T(\mathbf{p})-T(\mathbf{p}-\mathbf{k}%
)+\omega_{\mathbf{k}}]^{2}}\},\nonumber\\
\Gamma_{S}^{\prime}(t,\beta;\mathbf{p})  &  =\int\limits_{0}^{t}\gamma
_{S}^{\prime}(\tau,\beta;\mathbf{p})d\tau=\sum\limits_{\mathbf{k}%
}|V_{\mathbf{k}}|^{2}\frac{B_{S}(\mathbf{p})-B_{S}(\mathbf{p}-\mathbf{k}%
)}{B_{S}(\mathbf{p})}\nonumber\\
&  \times\{(1+N_{\mathbf{k}})[e^{i[T(\mathbf{p}-\mathbf{k})-T(\mathbf{p}%
)+\omega_{\mathbf{k}}]t}-1]\frac{1-e^{-[T(\mathbf{p}-\mathbf{k})-T(\mathbf{p}%
)+\omega_{\mathbf{k}}]\beta}}{[T(\mathbf{p}-\mathbf{k})-T(\mathbf{p}%
)+\omega_{\mathbf{k}}]^{2}}\nonumber\\
&  +N_{\mathbf{k}}[e^{i[T(\mathbf{p}-\mathbf{k})-T(\mathbf{p})-\omega
_{\mathbf{k}}]t}-1]\frac{1-e^{-[T(\mathbf{p}-\mathbf{k})-T(\mathbf{p}%
)-\omega_{\mathbf{k}}]\beta}}{[T(\mathbf{p}-\mathbf{k})-T(\mathbf{p}%
)-\omega_{\mathbf{k}}]^{2}}\},\nonumber\\
R_{S}(0,\beta;\mathbf{p})  &  =\rho_{\Sigma}(\beta)<\mathbf{p|}\rho_{S}%
(\beta)A_{S}(0)|\mathbf{p}>, \label{33}%
\end{align}
If the subsystem operator $A_{S}(0)$ depends only on the electron momentum
$\mathbf{p}$, then the initial value
\begin{equation}
R_{S}(0,\beta;\mathbf{p})=\rho_{\Sigma}(\beta)A_{S}(\mathbf{p})\rho_{S}%
(\beta;\mathbf{p}), \label{34}%
\end{equation}
where $A_{S}(\mathbf{p})=<\mathbf{p}|A_{S}(0)|\mathbf{p}>$, $\rho_{S}%
(\beta;\mathbf{p})=<\mathbf{p}|\rho_{S}(\beta)|\mathbf{p}>$.

Finally, the considered correlation function for electron momentums in the
second order in the small electron-boson field (thermostat) interaction is%
\begin{equation}
<A_{S}(t)B_{S}(0)>=\int d\mathbf{p}\rho_{S}(\beta;\mathbf{p})A_{S}%
(\mathbf{p})B_{S}(\mathbf{p})\exp\{-[\Gamma_{S}(t,\beta;\mathbf{p})+\Gamma
_{S}^{\prime}(t,\beta;\mathbf{p})]\}, \label{35}%
\end{equation}
where the relaxation functions $\Gamma_{S}(t,\beta;\mathbf{p})$ and
$\Gamma_{S}^{\prime}(t,\beta;\mathbf{p})$ are given by (\ref{33}) and the
function $\Gamma_{S}^{\prime}(t,\beta;\mathbf{p})$ defines the contribution of
initial correlations. Thus, we have obtained the solution for the electron
momentum-momentum correlation function in the second order in the
electron-phonon interaction, which shows, that generally, the damping process
is affected by initial correlations because $\Gamma_{S}^{\prime}%
(t,\beta;\mathbf{p})$ depends on time.

Although the obtained result, given by Eq. (\ref{32}), is valid for the whole
time range, there are some characteristic times in the problem under
consideration. When the subsystem-boson field interaction is small, we can
introduce the following time hierachy%
\begin{equation}
\tau_{rel}>>t_{0}. \label{36}%
\end{equation}
Here $\tau_{rel\text{ }}$is of the order of the electron relaxation time due
to the electron-bath interaction, $t_{0}=\max(t_{S},t_{\Sigma})$, where
$t_{S}$ is characteristic time of electron's collisions (e.g., $t_{S}%
\thicksim1/k_{B}T$) and $t_{\Sigma}$ is the characteristic time of correlation
of the thermostat fluctuations. It is instructive to consider the important
large timescale $t\thicksim\tau_{rel}>>t_{0}$ (in fact, it is possible to take
the limit $t\rightarrow\infty$). In this case, the relaxation functions
(\ref{33}) acquire more simple form%
\begin{align}
\Gamma_{S}(t,\beta;\mathbf{p})  &  =\tau_{rel}^{-1}(\beta,\mathbf{p}%
)\left\vert t\right\vert ,\Gamma_{S}^{\prime}(t,\beta;\mathbf{p})=i\frac{1}%
{2}\tau_{rel}^{-1}(\beta,\mathbf{p})\beta,\nonumber\\
\tau_{rel}^{-1}(\beta,\mathbf{p})  &  =2\pi\sum\limits_{\mathbf{k}%
}|V_{\mathbf{k}}|^{2}\frac{B_{S}(\mathbf{p})-B_{S}(\mathbf{p}-\mathbf{k}%
)}{B_{S}(\mathbf{p})}\nonumber\\
&  \times\{(1+N_{\mathbf{k}})\delta\lbrack T(\mathbf{p})-T(\mathbf{p}%
-\mathbf{k})-\omega_{\mathbf{k}}]+N_{k}\delta\lbrack T(\mathbf{p}%
)-T(\mathbf{p}-\mathbf{k})+\omega_{\mathbf{k}}]\},\nonumber\\
t  &  >>t_{0}. \label{37}%
\end{align}
To obtain (\ref{37}), we have used the definitions%
\begin{align}
\lim_{t\rightarrow\infty}\frac{\sin\omega t}{\omega}  &  =\pi\delta
(\omega),\nonumber\\
\lim_{t\rightarrow\infty}\frac{1-\cos\omega t}{\omega^{2}}  &  =\pi
\delta(\omega)\left\vert t\right\vert . \label{38}%
\end{align}
Note, that the relaxation function $\Gamma_{S}^{\prime}(t,\beta;\mathbf{p})$
ceases to depend on time. It means that on the large timescale the initial
correlations do not influence the damping of the correlation function in time.
Thus, on the timescale $t\thicksim\tau_{rel}>>t_{0}$, the correlation function
(\ref{35}) is%
\begin{align}
&  <A_{S}(t)B_{S}(0)>=\int d\mathbf{p}\rho_{S}(\beta;\mathbf{p})A_{S}%
(\mathbf{p})B_{S}(\mathbf{p})\exp[-\frac{i}{2}\tau_{rel}^{-1}(\beta
,\mathbf{p})\beta]\exp[-\tau_{rel}^{-1}(\beta,\mathbf{p})\left\vert
t\right\vert ],\nonumber\\
t  &  \thicksim\tau_{rel}>>t_{0}. \label{39}%
\end{align}
One can see, that on the large timescale the considered correlation function
damps exponentially with time. In addition, there is the factor $\exp
[-\frac{i}{2}\tau_{rel}^{-1}(\beta,\mathbf{p})\beta]$ describing the
contribution of initial correlations. This factor is absent in the paper
\cite{Los 1984}, where the consideration is restricted by the condition
$\tau_{rel}^{-1}\beta<<1$, when this factor is approximately equal to unity.
In the given paper, we take into consideration the initial correlations and
are able to calculate the corresponding corrections to the earlier obtained results.

\section{Polaron mobility}

According to the Kubo theory of linear response, the dissipative part of the
conductivity tensor is given by the following formula%
\begin{align}
\operatorname{Re}\sigma_{\mu\nu}(\omega)  &  =\frac{1}{E_{\beta}(\omega)}%
\int\limits_{0}^{\infty}\psi_{\mu\nu}(t)\cos(\omega t)dt,\nonumber\\
E_{\beta}(\omega)  &  =\frac{\omega}{2}\coth\frac{\beta\omega}{2},\nonumber\\
\psi_{\mu\nu}(t)  &  =\frac{1}{2}<j_{\mu}(t)j_{\upsilon}(0)+j_{\nu}(0)j_{\mu
}(t)>. \label{40}%
\end{align}
Here, $j_{\mu}(t)=e^{iHt}j_{\mu}(0)e^{-iHt}$ is the $\mu$-component of the
current operator in the Heisenberg representation, $H$ is the Hamiltonian in
the absence of an electric field, i.e. the Hamiltonian given by Eqs.
(\ref{17}) and (\ref{18}). Note, that $<j_{\nu}(0)j_{\mu}(t)>=<j_{\nu
}(-t)j_{\mu}(0)>$.

Let us consider the Fr\"{o}hlich polaron in the unit system with
$\hbar=m^{\ast}=\omega_{0}=1$, where $m^{\ast}$ is an electron effective mass
and $\omega_{0}$ is the crystal optical mode frequency. Then,
\begin{align}
H_{S}  &  =T(\mathbf{p})=\frac{\mathbf{p}^{2}}{2},H_{\Sigma}=\sum
\limits_{\mathbf{k}}b_{\mathbf{k}}^{+}b_{\mathbf{k}},H_{i}=\sum
\limits_{\mathbf{k}}V_{\mathbf{k}}e^{i\mathbf{kr}}b_{\mathbf{k}}%
+h.c.,\nonumber\\
V_{\mathbf{k}}  &  =2^{3/4}\pi^{1/2}\alpha^{1/2}/V^{1/2}k, \label{41}%
\end{align}
where $\alpha$ is the dimensionless constant of the electron-phonon
interaction and $V$ is the system's volume. Now, the above obtained results
for the correlation function can be directly applied to the calculation of the
polar crystal conductivity and the Fr\"{o}hlich polaron mobility. Since
$j_{\mu}(t)=ev_{\mu}(t)$ ($e$ is the electron charge, $v_{\mu}(t)$ is the
velocity operator), we need the velocity-velocity correlation function
$<v_{\mu}(t)v_{\nu}(0)>$, which can be calculated from Eq. (\ref{35}) for
arbitrary timescale. In particular, using Eq. (\ref{39}), we get for large
timescale $t>>t_{0}$
\begin{equation}
<v_{\mu}(t)v_{\nu}(0)>=\int d\mathbf{p}\rho_{S}(\beta;\mathbf{p})v_{\mu
}(\mathbf{p})v_{\upsilon}(\mathbf{p})\exp[-\frac{i}{2}\tau_{rel}^{-1}%
(\beta,\mathbf{p})\beta]\exp[-\tau_{rel}^{-1}(\beta,\mathbf{p})\left\vert
t\right\vert ],t>>t_{0}, \label{42}%
\end{equation}
where $\tau_{rel}(\beta,\mathbf{p})$ now is determined by (\ref{37}) with
$B_{S}(\mathbf{p})=v_{\upsilon}(\mathbf{p})=\partial T(\mathbf{p})/\partial
p_{\nu}$, and $V_{\mathbf{k}}$ and $T(\mathbf{p})$ are given by (\ref{41}).
This formula can be used for calculation of the low frequency conductivity
(\ref{40}), when $\omega<<t_{0}^{-1}$.

To simplify the consideration, let us consider the isotropic case, when
$\sigma_{xx}=\sigma_{yy}=\sigma_{zz}=\sigma$ (non-diagonal components are
equal to zero). Then, the relaxation frequency (\ref{37}) for the correlation
function $<v_{z}(t)v_{z}(0)>$ after integration over $k$ acquires the form
(see also \cite{Los 1984})%
\begin{align}
\lbrack\tau_{rel}^{-1}(\beta,\mathbf{p})]_{z}  &  =\Gamma_{z}(\beta
,\mathbf{p})=\frac{\alpha}{\sqrt{2}\pi p_{z}}[(1+N_{0})%
%TCIMACRO{\dint \limits_{(p\cos\theta_{1}>\sqrt{2})}}%
%BeginExpansion
{\displaystyle\int\limits_{(p\cos\theta_{1}>\sqrt{2})}}
%EndExpansion
d\Omega\cos\theta\frac{2p\cos\theta_{1}}{\sqrt{p^{2}\cos^{2}\theta_{1}-2}%
}\nonumber\\
&  +N_{0}%
%TCIMACRO{\dint }%
%BeginExpansion
{\displaystyle\int}
%EndExpansion
d\Omega\cos\theta\frac{p\cos\theta_{1}}{\sqrt{p^{2}\cos^{2}\theta_{1}+2}}].
\label{43}%
\end{align}
Here, $N_{0}=(e^{\beta}-1)^{-1}$, $d\Omega=d\varphi\sin\theta d\theta$, the
angles $\varphi$ and $\theta$ define the direction of the wave vector
$\mathbf{k}$, and $\theta_{1\text{ }}$is the angle between $\mathbf{p}$ and
$\mathbf{k}$, which is related to $\varphi$ and $\theta$ as
\begin{equation}
p\cos\theta_{1}=\sin\theta\cos\varphi p_{x}+\sin\theta\sin\varphi p_{y}%
+\cos\theta p_{z}. \label{44}%
\end{equation}
The area of integration in the first integral is defined by the condition of
the integrand positive value.

It is interesting to consider the low temperature case $\beta>>1$
($k_{B}T<<\omega_{0}$), which was under debate for more than thirty years (see
Introduction and \cite{Devreese arXiv}). As was mentioned, the correct
low-temperature polaron mobility, which agrees with the experiment, was
obtained in \cite{Los 1984}. In the present paper, we are able to consider the
influence of initial correlations on the low-temperature polaron mobility. It
is seen from (\ref{35}) and (\ref{41}), that at low temperature $\beta>>1$,
the main contribution to correlation function comes from the area of small
electron momenta $p^{2}<<2$. Thus, in this case, the first term in the
relaxation frequency (\ref{43}), which corresponds to the emitting of phonon
by electron, vanishes due to impossibility to satisfy the energy and momentum
conservation law. In the second term of (\ref{43}), we can omit $p^{2}\cos
^{2}\theta_{1}$ in the denominator. As a result, this term, which describes
the process of the phonon adsorption by an electron, after integration reduces
to%
\begin{equation}
\Gamma_{z}(\beta,\mathbf{p})=\Gamma^{0}=\frac{2}{3}\alpha N_{0},p^{2}<<2.
\label{45}%
\end{equation}
Thus, at small momenta, the relaxation frequency does not depend on
$\mathbf{p}$.

Now, we can calculate from (\ref{40}) the low-temperature electron
conductivity of the Fr\"{o}hlich polaron. To do that, we also need the
correlation function $<v_{z}(0)v_{z}(t)>=<v_{z}(-t)v_{z}(0)>$. It is not
difficult to see from (\ref{33}) and (\ref{38}), that this correlation
function at $t>>t_{0}$ is given by (\ref{42}) with the substitution
$\beta\rightarrow-\beta$. Substituting (\ref{42}) with the relaxation
frequency (\ref{45}) and the corresponding expression for $<v_{z}%
(-t)v_{z}(0)>$ into (\ref{40}), we obtain the following low frequency
conductivity of the Fr\"{o}hlich polaron
\begin{equation}
\sigma(\omega)=ne^{2}<v_{z}^{2}>\beta\frac{\Gamma^{0}}{\omega^{2}+(\Gamma
^{0})^{2}}\cos(\Gamma^{0}\beta),\omega<<\beta^{-1}<<1, \label{46}%
\end{equation}
where $n$ is the electron concentration, $<v_{z}^{2}>=\int d\mathbf{p}\rho
_{S}(\beta;\mathbf{p})v_{z}^{2}(\mathbf{p})$.

Thus, in the considered range of the parameters, we have the Drude formula for
conductivity. The low-temperature polaron mobility, as it follows from
(\ref{45}) and (\ref{46}), is
\begin{align}
\mu &  =\sigma(0)/ne=\mu_{0}-\Delta\mu,\beta>>1,\nonumber\\
\mu_{0}  &  =\frac{3e}{2\alpha}e^{\beta},\Delta\mu=\frac{e}{3}\alpha\beta
^{2}e^{-\beta}, \label{47}%
\end{align}
where we used, that with a good accuracy $<v_{z}^{2}>=\beta^{-1}$ (in the
adopted unit system), and expanded $\cos(\Gamma^{0}\beta)$ up to the second
power in the small term $\Gamma^{0}\beta<<1$.

The obtained result shows, that accounting for initial correlations results in
the very small correction $\Delta\mu<<\mu_{0}$ to the expression for the
low-temperature polaron mobility $\mu_{0}$ found in \cite{Los 1984} and
confirmed later in \cite{Sels and Brosens 2014} (see also discussion in
\cite{Devreese arXiv}). Anyway, this correction is important, since we have
not used any unrealistic approximation for the initial state of the system
under consideration. However, it should be kept in mind, that the result
(\ref{47}) is valid for $\omega=0$, i.e. for a large timescale ($t\rightarrow
\infty$) when, as was mentioned earlier, the initial correlations do not
influence the time relaxation process. For shorter times (larger frequencies
$\omega$) the initial correlations can influence the damping of the
correlation function in time (see Eq. (\ref{35})). Thus, it is of principal
importance to have a comprehensive approach to calculation of contribution of
initial correlations to the evolution of the subsystem, which has been
suggested in this paper.

\section{Conclusion}

We have derived the exact equation (\ref{13}), which defines the evolution in
time of the two-time correlation function (\ref{2}) of a subsystem interacting
with a boson field (a heat bath) and accounts for initial correlations. The
obtained equation provides the regular procedure for including into
consideration the influence of initial correlations (ignored as a rule) on the
evolution process. Initial correlations enter the kernel govering the
correlation function evolution on an equaI footing with the terms defining the
collisions of a subsystem with the heat bath excitations. In the second order
of the particle-bath interaction expansion of the kernel governing the
evolution of the electron momentum-momentum correlation function, we have
obtained the solution to this correlation function (Eq. (\ref{35})) on an
arbitrary timescale. It shows that generally the initial correlations
influence the evolution in time process. However, on a large timescale, this
influence vanishes (in accordance with Bogoliubov's \ principal of initial
cortrelations) and the initial correlations contribute only to the amplitude
of the exponential damping of the corresponding dynamical \ variable (see Eq.
(\ref{39})). As an application, the low-temperature Fr\"{o}hlich polaron
mobility is considered. We found a small correction to our earlier result
\cite{Los 1984} which shows that initial correlations slightly reduce the
low-temperature polaron mobility.

\end{document}